**Title:** Metrology for Inductive Charging of Electric Vehicles (MICEV)

Authors: Mauro Zucca ; Oriano Bottauscio ; Stuart Harmon ; Roberta Guilizzoni ; Florian Schilling ; Matthias Schmidt ; Peter Ankarson ; Tobias Bergsten ; Kari Tammi ; Panu Sainio ; J. Bruna Romero ; E. Laporta Puyal ; Lionel Pichon ; Fabio Freschi ; Vincenzo Cirimele ; Pavol Bauer ; Jianning Dong ; Antonio Maffucci ; Salvatore Ventre ; Nicola Femia ; Giulia Di Capua ; Niels Kuster ; Ilaria Liorni

**Acknowledgement:** The results presented in this paper are developed in the framework of the 16ENG08 MICEV Project. The latter received funding from the EMPIR programme co-financed by the Participating States and from the European Union's Horizon 2020 research and innovation programme.

# Metrology for Inductive Charging of Electric Vehicles (MICEV)

Mauro Zucca, Oriano Bottauscio
*Istituto Nazionale di Ricerca Metrologica - INRiM*
Torino, Italy
m.zucca@inrim.it
o.bottauscio@inrim.it

Peter Ankarson, Tobias Bergsten
*Research Institutes of Sweden RISE*
Boras, Sweden
peter.ankarson@ri.se
tobias.bergsten@ri.se

Lionel Pichon
*Laboratoire Génie électrique et électronique de Paris -GeePs*
Paris, France
lionel.pichon@supelec.fr

Antonio Maffucci, Salvatore Ventre
*Università degli studi di Cassino e del Lazio Meridionale - UNICAS*
Cassino, Italy
maffucci@unicas.it
ventre@unicas.it

Stuart Harmon, Roberta Guilizzoni
*National Physical Laboratory NPL*
Teddington, UK
stuart.harmon@npl.co.uk
roberta.guilizzoni@npl.co.uk

Tammi Kari, Panu Sainio
*Aalto University Aalto*
Espoo, Finland
kari.tammi@aalto.fi
panu.sainio@aalto.fi

Fabio Freschi, Vincenzo Cirimele
*Politecnico di Torino POLITO*
Torino, Italy
fabio.freschi@polito.it
vincenzo.cirimele@polito.it

Nicola Femia, Giulia Di Capua
*Università degli studi di salerno - UNISA*
Fisciano (SA), Italy
femia@unisa.it
gdicapua@unisa.it

Florian Schilling, Matthias Schmidt
*Physikalisch-Technische Bundesanstalt PTB*
Braunschweig, Germany
florian.schilling@ptb.de
Matthias.Schmidt@ptb.de

J. Bruna Romero, E. Laporta Puyal
*Research Centre for Energy Resources and Consumption - Fundación CIRCE*
Zaragoza, Spain
jbruna@fcirce.es
elaporta@fcirce.es

Pavol Bauer, Jianning Dong
*Delft University of Technology TU Delft*
Delft, Netherlands
p.bauer@tudelft.nl
j.dong-4@tudelft.nl

Niels Kuster, Ilaria Liorni
*Schmid & Partner Engineering AG - SPEAG*
Zürich, Switzerland
nk@itis.ethz.ch
liorni@itis.swiss

*Abstract*— The European Union funded project MICEV aims at improving the traceability of electrical and magnetic measurement at charging stations and to better assess the safety of this technology with respect to human exposure. The paper describes some limits of the instrumentation used for electrical measurements in the charging stations, and briefly presents two new calibration facilities for magnetic field meters and electric power meters. Modeling approaches for the efficiency and human exposure assessment are proposed. In the latter case, electromagnetic computational codes have been combined with dosimetric computational codes making use of highly detailed human anatomical phantoms in order to establish human exposure modeling real charging stations. Detailed results are presented for light vehicles where, according to our calculations, the concern towards human exposure is limited. Currently, the project has reached half way point (about 18 months) and will end in August 2020.

I. INTRODUCTION

Inductive power transfer (IPT) for electric vehicles charging offers undoubted advantages, despite a slight reduction in the charging energy efficiency. The advantages are considerable: IPT realizes a real autonomous driving and does not require actions by the driver to recharge, except parking in appropriate areas or driving in specific lanes. According to [1], for light vehicles the minimum target efficiency of inductive charging is 85% and at least 80% in a displaced position (incorrect parking). Researchers are making efforts to increase this efficiency and bring it to levels similar to conductive charging (92% - 95%) [2]. Furthermore, in order for this technology to be publicly accepted, clear, effective and reliable evidence that it is safe for humans should be provided.

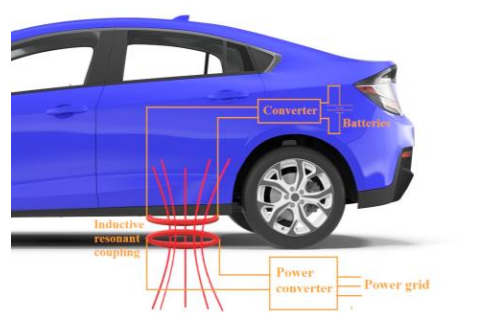

Fig. 1 – Scheme of a static charging system.

The first measurement challenge is how to increase accuracy and traceability in the measurement of the power transfer efficiency. The power measurement in these systems takes place in a frequency range 20 kHz to 150 kHz. The total harmonic distortion (THD) can reach 2% for the current and more than 45% for the voltage in the resonant circuit. In this frequency range a power analyzer is normally used together with current transducers, and sometimes with voltage transducers (in the first instance to preserve the power analyzer). As we will see later, this can lead to a measurement uncertainty that can be about 10% at 100 kHz.

The second measurement challenge relates to the assessment of human exposure due to electromagnetic fields. From this point of view, scientific literature is proliferating, [3-6], even though many studies, particularly with real car bodies, are still missing. Even from the regulatory point of view, there are written standards in draft, such as [7], but this specific regulatory activity is on-going and requires more input from research. The following sections illustrate how this project is contributing to the above stated challenges.

More specifically, the scientific objectives of the MICEV project are the following:

- To develop and calibrate with high accuracy (relative uncertainty in the DC circuit of $10^{-3}$) a power measurement unit for static wireless power transfer for on-board measurement and for the efficiency assessment of the IPT.

- To develop measurement protocols for the assessment of human exposure to electromagnetic fields adopting compliance with the limits indicated by [8-9].

To achieve this, the partners have been carrying out an extensive simulation survey based on advanced electromagnetic models. Moreover, to increase accuracy in the assessment of the human exposure an additional objective of the project is:

- To develop a measurement systems for traceable calibration of magnetic field meters up to 150 kHz and up to 100 µT, including field gradients with both sinusoidal and non-sinusoidal waveforms. The target expanded uncertainty for the system is 5 %.

The project also aims at a metrological analysis of dynamic inductive charging, which will be discussed in future papers.

The results here presented have been developed in the framework of the EMPIR 16ENG08 MICEV Project. The EMPIR initiative is co-funded by the European Union's Horizon 2020 research and innovation programme and the EMPIR participating States.

## II. POWER EFFICIENCY

*A. A charging station*

A charging station includes a connection to the electricity grid, where an isolation transformer can sometimes be inserted. Downstream, an AC/AC converter transforms the power frequency voltage to a voltage at the transmission frequency, being from 20 kHz - 30 kHz for heavy vehicles, up to about 85 kHz (81.38 kHz ≤ f ≤ 90 kHz) for light vehicles (power ≤ 20 kW, typically 7.7 kW).

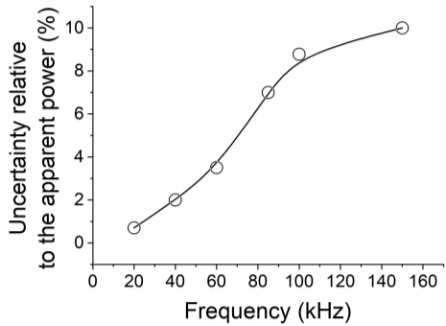

Fig. 2 – Power measurement uncertainty for on board measurements using a power analyzer and zero flux current transducers.

The converter supplies an RC resonant circuit, where the inductance is mainly due to the transmission coil. The latter is coupled to another coil on board the vehicle. This latter coil is also connected to a RC circuit, which resonates at the same frequency as the ground coil. The signal collected by the on-board coil is then rectified and eventually filtered by a converter system, and it charges the car batteries. Fig. 1 summarizes the general scheme.

*B. Measuring the overall charging efficiency*

The measurement of efficiency can be given as the ratio of the power measured on board and the power absorbed from the electrical grid. For industrial measurements, an uncertainty of the order of 1% relative can be considered good. In this framework, grid side measurement does not usually give accuracy problems within the accuracy target mentioned above. The measurement on board may be more problematic. To evaluate this effect, the project partners prepared a state-of-the-art commercial power analyzer fitted with commercial high quality closed loop current transducers using a zero flux detector. The results of the calibration performed at RISE are shown in Fig. 2, where the expanded uncertainty reaches 10% at 150 kHz and is higher than 6% at 85 kHz.

These results become significantly worse if voltage transducers are also used, and a specific study is ongoing on this topic. The voltage transducers introduce additional ratio and phase errors. A typical phase error response of a good commercial transducer is shown in Fig. 3.

On the contrary, for network frequency measurements, there are no particular problems, since, with similar instrumentation, the relative error is below 200 ppm.

One way to overcome the above errors, in particular for the measurements in the resonant circuit, is to use coaxial shunts as current transducers, and to correct the ratio and phase errors of the voltage transducers. Guidance on how to measure the phase error of voltage transducers is provided by INRIM in [10].

*C. The validation of the measurement system*

When an on-board power measurement system is finally well thought out, it must be calibrated.

As for on-board power measurements, these can be performed in the resonant circuit. In this case the actual power transferred on board is measured and the power is associated with a rectangular voltage waveform and a nearly sinusoidal current, in which usually two small third and fifth harmonic contributions are present (Fig. 4). A simpler and more accurate method is required to measure the battery charging power. In this second case voltages and currents are a continuous signals with the overlap of a ripple, both for the voltage and for the current, whose frequency is linked to the switching frequency of the power converters (Fig 5). Therefore, also in this second case, it is necessary to measure a high-frequency power component whose entity is the subject of an ongoing investigation within MICEV.

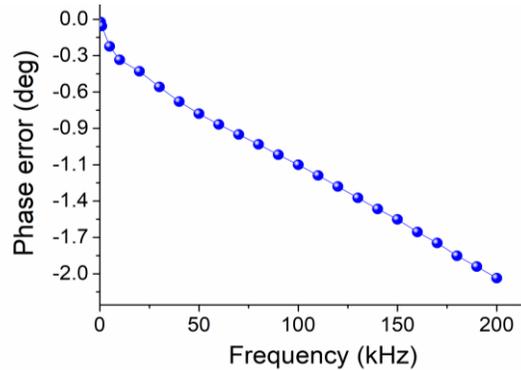

Fig. 3 – Phase error versus frequency of a high-level commercial voltage transducer.

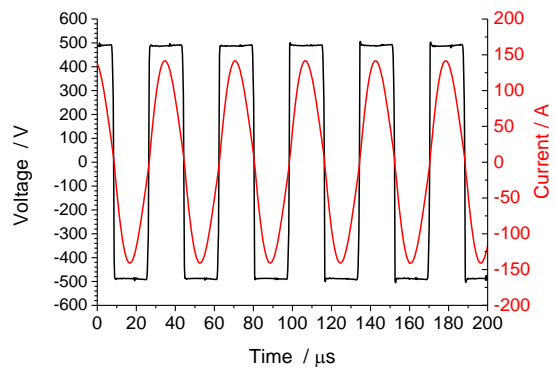

Fig. 4 – Measured voltage and current waveforms in the resonant circuit.

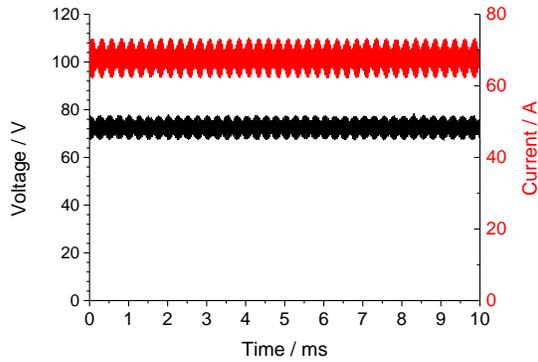

Fig. 5 – Measured voltage and curr ent waveforms at the batteries.

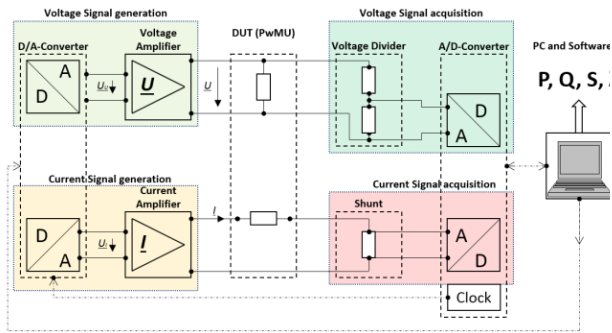

Fig. 6 – Reference measurement system based on an AC phantom power.

For both of the cases mentioned above, it is a question of measuring a power with a frequency of tens of kilohertz, partly distorted, and possibly with a superposed DC component. The accuracy of the measurement is influenced both by the angle and the ratio errors. A phase error of 2 degrees at 100 kHz (55 ns) when the phase shift between voltage and current is 45°, leads to an error in the active power measurement of 3.5%. Errors of such kind must be evaluated and, when possible, compensated. Moreover, the measurement systems should be validated by calibration.

Currently, there are no calibration facilities designed for this type of application, but within the MICEV project one was built at PTB. Fig. 6 illustrates the configuration of the phantom power setup, where two independent circuits, to which a reduced power is associated, contribute to simulate a greater power for the reference measurement system, which take part in the same measurements (voltage, current and power) as the system to be characterized. The built-up generation systems can synthesize realistic test signals in the laboratory environment and generate a phantom power from the synthesized signals with a LabVIEW™ software. This latter software tool induces the signal sequences to generate the current signal and the voltage signal via a connected D/A converter, and it computes the parameters of the resulting signal sequences at the amplifiers output. The voltage measurements, performed in parallel for system and DUT, and current measurements done in series are collected to the software for the electrical voltage, current and power calculation.

### D. Modeling the efficiency

In the frame of MICEV project, a proper modelling activity is carried out with the main goal of analyzing the sensitivity of the main features of the system (including power efficiency) with respect to the design parameters, to provide suitable design guidelines.

The model of the overall system may be conceptually subdivided into two blocks: (i) the coupling system, including the coil pairs with the associated conducting (e.g., chassis) and magnetic (e.g., ferrites) components; (ii) the power electronics, including transformers, converters, controllers and so on. The system-level analysis of the power efficiency is carried out on an equivalent circuit model in which the coupling system is modeled as a real coupled inductor pair, *i.e.* with self and mutual inductances and equivalent resistances accounting for the losses in the coils and in the conducting regions (chassis). Looking at the coupling system, the most relevant parameter affecting the overall efficiency is the mutual inductance, $M$. Therefore, a sensitivity analysis must be primarily addressed to check the variation of $M$ with respect to the system parameters. To this purpose, we have analyzed the sensitivity of $M$ with respect to the geometrical mismatch between the nominal position of the coils and the real one, assuming the nominal parameters of the 85 kHz-IPT system analyzed in MICEV project. The two coils lay on the *z-y* plane (see Fig.7), and the nominal position is obtained by putting $\Delta z = \Delta y = 0$ $cm$ and $\alpha = 0$ An additional variability may be included by assuming a vertical distance along *x*-axis ($\Delta x$) that can differ from its nominal value. Figure 8 shows the behavior of $M$ versus a horizontal misalignment $\Delta z$, for different values of a rotation angle $\alpha$.

The complete model of the coupling coils include an aluminum structure (which simulates the vehicle chassis) and some ferrite parts, hence the extraction of the equivalent parameters requires a full 3D numerical solution of a Magneto-Quasi-Static problem. The cost of such a simulation may easily become unaffordable when a large number of case-studies are to be mapped, as in this sensitivity analysis. To overcome this problem, a behavioral model of mutual inductance $M$ has been derived by using the Genetic Programming algorithm. In this approach, the population is composed of models and their evolution is dictated by classical genetic operations, such as selection, cross-over, mutation. This approach has been here applied to model the dependence of $M$ on $\Delta x$, $\Delta z$, and $\alpha$. Specifically, $M_{behav}(\Delta x, \Delta z, \mathbf{p}(\alpha))$ is given as a function of $\Delta x$ and $\Delta z$:

$$M_{behav} = -p_0 \Delta z^{1.5} \exp(-p_1 \Delta x) + p_2 \exp(-p_3 \Delta x) \quad (1)$$

and is parametrized with respect to $\alpha$, being $\mathbf{p}(\alpha)$ a vector of model coefficients dependent on $\alpha$:

$$p_i = a_{i,0} \cos(2\alpha) + a_{i,1}, \quad i = 0,1,2,3 \quad (2)$$

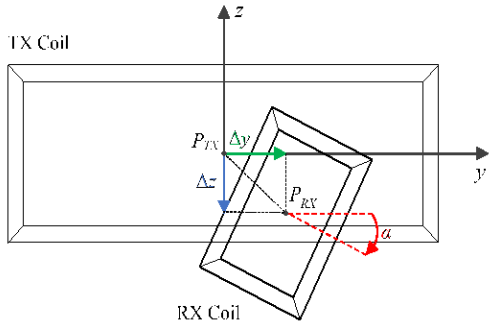

Fig. 7 – The two coil system in a generic, misaligned position.

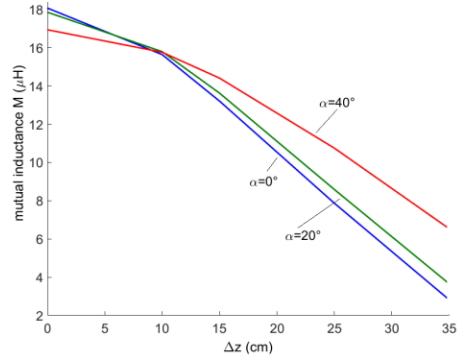

Fig. 8 – Sensitivity analysis of the mutual inductance for the coil pair in Fig. 7, with respect to an in-plane shift (Δz) and a rotation (α) of the coil reciprocal position .

Figure 9 shows that the percent error introduced by the behavioral model in the 300 test conditions of the training data set (300 different values of coils misalignment in terms of $\Delta x$, $\Delta z$ and $\alpha$) is less than 6%.

### III. HUMAN EXPOSURE MEASUREMENTS AND MODELS

For the evaluation of the human exposure to electromagnetic fields at charging stations, one can operate in two ways: through verification measurements on commercially manufactured IPT systems, or through models created before or after the realization of such systems.

*A. Verification measurements*

The measurement of magnetic induction around the IPT systems requires the calibration of commercial field meters. Traceability of these measurements is limited, since the calibration requirement for high magnetic induction values is mainly required at power frequencies, 50/60 Hz up to a few kilohertz. For field generating coil systems, the maximum amplitude decreases with increasing frequency and currently at 100 kHz the maximum amplitude for traceability is 25 µT [11]. Within the MICEV project, traceable measurement capabilities in Europe are extended, with a new system developed at NPL, currently being validated (Fig. 10).

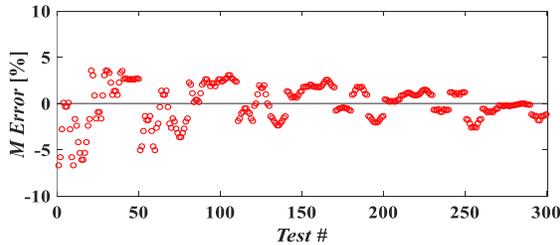

Fig. 9 – Percent error introduced in the mutual inductance prediction by the use of the behavioral model (1)-(2).

The new calibration system is a Helmholtz coil configuration, consisting of two 150 mm radius coils made of Type 2 Litz wire, each with 4 turns. The initial DC coil constant (H/I, magnetic field strength to current ratio) was determined to be 18.62 A/m/A. For AC magnetic field meters, the AC frequency response is required and this was evaluated using a single turn search coil. Figure 11 shows the H/I vs frequency up to 150 kHz. Axial and radial variations were also measured and compared against modelled values as the uniformity of the field within the Helmholtz coils is a key component of the uncertainty evaluation.

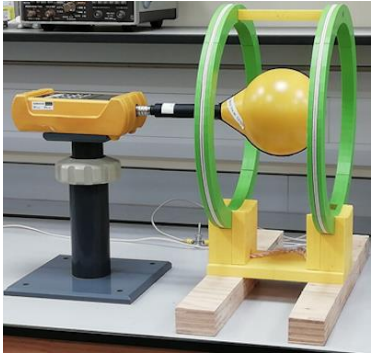 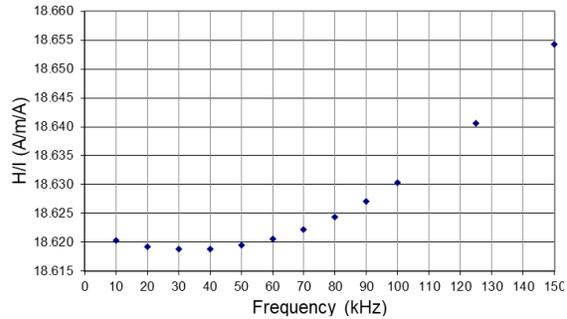

Fig. 10– New calibration facility for magnetic field meters at NPL.

Fig. 11 – Frequency response of new Helmholtz coil system designed and built at NPL.

The calibration facility will be validated through a tri-lateral comparison. The travelling standard (5 Hz to 250 kHz magnetic field Loop Sensor, 4 cm diameter, 51 turns) has been selected and purchased and its characterisation is in progress.

The project also aims to give traceability to gradient measurements for AC magnetic fields, which are also required for the assessment of exposure. On this purpose a new gradient field probe is being developed within the project and a new verification system will be characterized on the basis of this probe.

### B. Charging stations modeling approach

The modeling approach allows to evaluate the human exposure in different scenario of the charging system, according to various parameters such as misalignment, distance between the coils, position of the coils under the vehicle etc. Moreover, through bioelectromagnetic models, it is possible to calculate the actual levels of induced electric field in subjects with different morphology (infants, children and adults) and in real postures. The spatial distribution of the magnetic field is assumed as input for the bioelectromagnetic simulation (see Sect. C), thus the results provided by the preliminary electromagnetic simulation must be accurate and reliable. A preliminary computation was performed to assess the effectiveness of the geometrical mesh chosen by each partner across different numerical simulation software. This purpose was achieved using an axial symmetrical charging system shown in Fig. 12, where two coils are considered with a radius of 280 mm, with electric currents in quadrature equal to 500 Aturns for the transmitting coil and 1050 Aturns for the receiving coil.

A ferrite concentrator and an aluminium shield complete the system. The code validation has been achieved by comparing the results both with 2D and 3D codes, both commercial (Comsol, Opera 3D, Maxwell, Sim4Life, CST) and proprietary (Sally2D), until a satisfactory agreement was obtained, with a discrepancy lower than 6% on the peak values. Results are shown in Fig. 13.

After this preliminary step, two charging stations were modeled, one for light vehicles working at 85 kHz and one for heavy vehicles working at the frequency of ~25 kHz (minibuses) based on CIRCE's Victoria platform (Fig. 14). The IPT system of the light vehicle was modeled on the basis of the designs suggested by the J2954 SAE standard [1]. A WPT2/Z3 power class system defined in SAE J2954 was modeled, for a maximum power of 7.7 kVA. The system includes an aluminum shield and a ferrite concentrator. The magnetomotive force in the primary and secondary coils reach ~ 200 Aturns. The car body is a real one provided by a car manufacturer. The physical properties of the car body can vary as follows: relative permeability from 30 to 300, electrical conductivity from 2 MS/m to 7 MS/m. The same physical properties were used for the minibus, but in this case the maximum rated power is about 50 kVA.

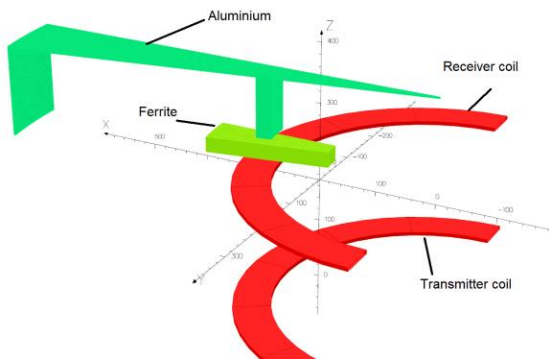

Fig. 12 – Double section of the axisymmetric charging system. Dimensions in mm.

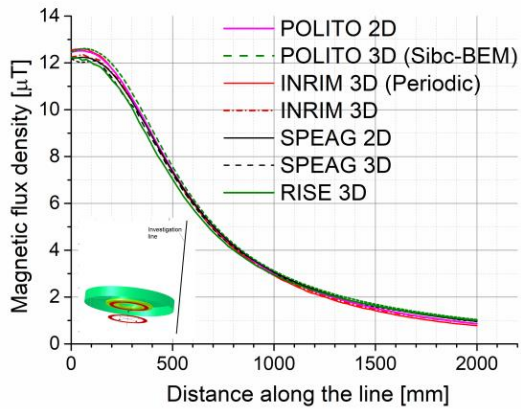

Fig. 13 – Results comparison along a 2 m z directed line of the numerical codes from the partners in order to assess the accuracy of the preliminary numerical computations (magnetic field distribution) for the bioeletromagnetic models.

The stray magnetic field generated around these systems is computed by a finite element electromagnetic solver, able to account for the shielding effect due to the presence of the car body, usually having magnetic and conductive properties. Results have been assessed comparing different commercial (Cobham Opera 3D [12] and the low frequency module of CST Studio Suite [13]) and research codes [14].

### C. Bioelectromagnetic simulations

The stray magnetic field generated in the vicinity of inductive charging systems is far from being uniform, therefore we have to refer to the basic restriction (i.e. induced electric field in the human body) in order to evaluate the compliance with ICNIRP Guidelines [8-9]. To perform this analysis, highly detailed anatomical human models and reliable electromagnetic tools for computing induced electric fields in tissues are needed.

The Project Consortium has adopted the selection of highly detailed human anatomical phantoms belonging to the Virtual Population (ViP) developed by IT'IS Foundation [15]. These phantoms are representative of an adult male (Duke) and a female (Ella), a child (Thelonious) and a baby (Charlie). All phantoms are posable (so their posture can be arranged to simulate the exposure in a more realistic way) and are segmented into more than 300 different tissues. Dielectric tissue properties were assigned according to the IT'IS Tissue Database [16], but the skin tissue whose conductivity was set to 0.1 S/m to account for the deeper granular tissue (i.e., the dermis).

In-silico computation of the spatial distribution of the induced electric field has been performed under the quasi-static approximation, by assuming that the magnetic field is not perturbed by the currents induced in the body. The spatial distribution of the magnetic field, generated by the IPT systems and computed as described in the previous section, has been assumed as input for the bioelectromagnetic simulation. The maximum of the induced E fields and current density are determined according to the current guidelines [ICNIRP 1998, ICNIRP 2010, IEEE 2005] in each tissue and organ. In case the specific averaging method is not enough to avoid unwanted artefacts, further filtering techniques are adopted to remove local numerical artefacts caused by the spatial discretization of the human body (voxel size and staircase effects) [17]. Significant postures of the human anatomical phantoms have been generated through the Poser tool available in the simulation software Sim4Life, in order to assess the exposure when the human anatomical phantoms are placed outside and inside the vehicles.

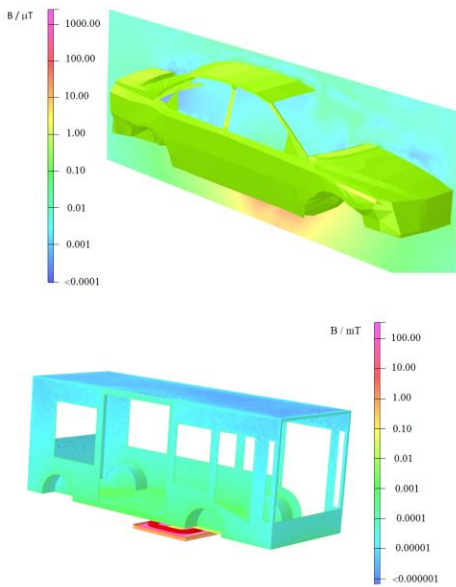

Fig. 14 – Complete modeling of charging stations with vehicles.

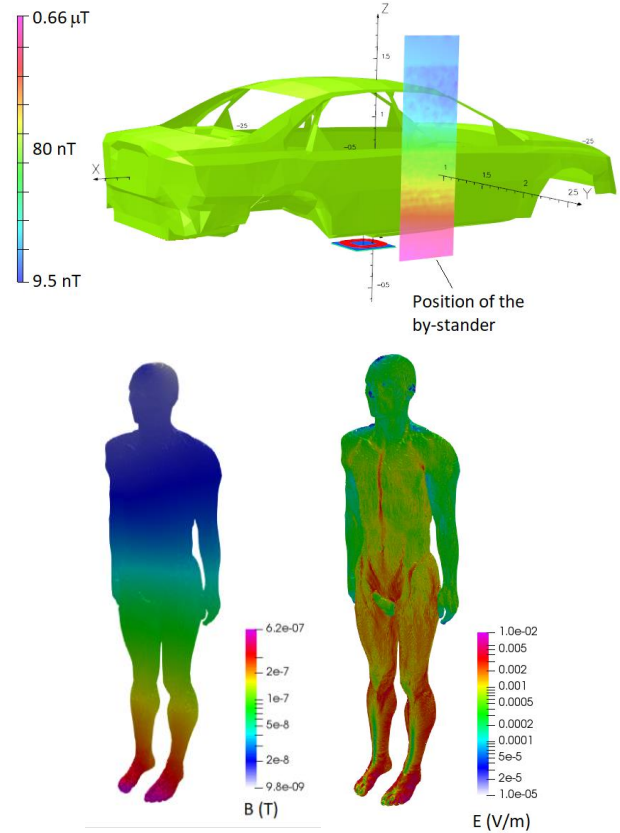

Fig. 15 – Example of exposure of a by-stander close to the car: (a) car body and position of the by-stander, (b) surface values of the stray magnetic flux density, (c) surface values of the induced E field.

An example of human exposure of a by-stander (Duke phantom) placed at a distance of about 0.2 meters from the car body of a sedan car is reported in Fig. 15. This example highlights how the levels of magnetic flux density (lower than 0.7 µT) and induced electric field in the human body (maximum of 10 mV/m) are considerably lower with respect to the ICNIRP 2010 reference level and basic restriction, which are 27 µT and 11.5 V/m, respectively, at 85 kHz.

## IV. Conclusions

The ongoing MICEV project has begun to contribute to the improvement of measurement in vehicle inductive charging systems. The accuracy limits of commercial measurement equipment for electrical measurements have been analysed. Two new specific facilities for the calibration of magnetic field meters and power meters have been achieved, and one for gradient field measurement is being developed. Such facilities extend measurement capabilities in Europe. Detailed dosimetric models and phantoms have been setup to reproduce the human exposure in real charging stations. For light vehicles, human exposure resulted to be considerably lower than basic restrictions. Regarding the exposure in case of heavy vehicles, the efficiency measurements and the dynamic charging, many activities are still in progress and will produce results in the near future.